\documentclass[12pt]{article}

\pdfoutput=1

\usepackage{cite,color}
\usepackage{mathrsfs,amsmath,amssymb,bm,fixmath}

\catcode`@=11

\let\@CITE=\@cite
\let\Cite=\cite
\def\@cite#1#2{{#1\if@tempswa , #2\fi}}

\def\bcite{\Cite}
\def\cite#1{[\Cite{#1}]}

\catcode`@=12

\evensidemargin=\oddsidemargin
\def\Eqn#1{Eq.\ (\ref{#1})}
\def\Eqs#1#2{Eqs.\ (\ref{#1}) and (\ref{#2})}
\def\3Eqs#1#2#3{Eqs.\ (\ref{#1}), (\ref{#2}) and (\ref{#3})}

\def\textcite#1{Ref.\ \bcite{#1}}

\def\del{\partial}
\def\v0{ \xleftarrow{\quad v=0 \quad}}

\title{\bf Reply to ``Comment on `Covariant formulation of
  electrodynamics in isotropic media'\,''}

\author{\bf Palash B. Pal \\
Physics Department, University of Calcutta,\\ 92 APC Road, Calcutta
700009, India}

\date{}

\begin{document}

\maketitle

\begin{abstract}

This note contains my response to the comment written by J. Franklin
on my paper ``Covariant formulation of electrodynamics in isotropic
media''.
  
\end{abstract}

\bigskip

Jerrold Franklin \cite{comment} has raised an important question about
my paper ``{\em Covariant formulation of electrodynamics in isotropic
  media}'' \cite{mypaper}.  It concerns a point that should have been
clarified in my original paper.  I regret that I did not do it, and
thank Franklin for opening up the discussion.  As I will show, I do
not agree with his final conclusion.

The crux of Franklin's argument lies in the question of whether the
electromanetic scalar and vector potential constitute a 4-vector
$A^\mu$.  Denoting the combination $\{\phi, \vec A\}$ by $A^\mu$, the
Maxwell equations in vacuum can be written in the
form\footnote{\textcite{comment} uses Gaussian units with $c=1$.  In
  keeping with the convention of my original paper \cite{mypaper}, I
  use the Heaviside-Lorenz units, and write the factors of $c$
  explicitly.}
\begin{eqnarray}
  \Box A^\mu = \frac1c J^\mu  \,,
  \label{boxA}
\end{eqnarray}
where
\begin{eqnarray}
  \Box = {1 \over c^2} \, {\del^2 \over \del t^2} - \vec\nabla^2 =
  \del_\mu \del^\mu \,,
\end{eqnarray}
and the Lorenz gauge has been employed for the potentials.  Franklin
argues that since $\Box$ is a Lorentz-invariant differential operator
and $J^\mu$ is a true 4-vector, \Eqn{boxA} implies that $A^\mu$ is a
4-vector in the vacuum.  I have a comment on this line of reasoning,
which I will present at the end of this note.  Right now, let us
accept Franklin's argument and proceed.

Within a medium, however, \Eqn{boxA} is modified to
\begin{subequations}
  \label{medeq}
\begin{eqnarray}
  \epsilon \left( {\epsilon\mu \over c^2} \del_t^2 - \vec\nabla^2
  \right) \phi &=& \rho \,, 
  \label{medeq1} \\ 
  \frac1\mu \left( {\epsilon\mu \over c^2} \del_t^2 - \vec\nabla^2
  \right) \vec A &=& \frac1c \vec J \,, 
  \label{medeq2}
\end{eqnarray}
\end{subequations}
where $\rho$ and $\vec J$ represent the external sources only.  
Franklin argues that, since the differential operator on the left side
of these equations is not Lorentz invariant, it does not follow that
the foursome $\{\phi,\vec A\}$ is a 4-vector.  This is where I
disagree.

Let us examine the left sides of \Eqn{medeq}.  I already argued in my
original paper \cite{mypaper} that $\epsilon$ and $\mu$ are
Lorentz-invariant quantities.  Franklin \cite{comment} also did not
contest this argument.  As regards the parenthesized differential
operator on the left side of \Eqn{medeq}, it should be remembered that
\Eqn{medeq} is valid {\em only} in the rest frame of the medium, and
therefore the covariance of the equations should not be judged from
\Eqn{medeq}.  The crucial question is whether we can write a covariant
set of equations which reduce to \Eqn{medeq} in the rest frame of the
medium.  In this pursuit, one needs to remember that the presence of a
medium presupposes a 4-vector $u^\mu = \{ \gamma, \gamma \vec v/c \}$,
which is the 4-velocity of the medium in the frame concerned.  This
allows us to write a relation of the form
\begin{subequations}
  \label{invops}
\begin{eqnarray}
  {1 \over c^2} \del_t^2 \v0 u^\mu u^\nu \del_\mu \del_\nu  \,,
\end{eqnarray}
which means that the expression on the left side can be seen as the
$v=0$ limit of the Lorentz-invariant expression written on the right
side.  In other words, the Lorentz-invariant expression on the right
side reduces to the expression on the left side in the rest frame of
the medium.  Similarly,
\begin{eqnarray}    
  \vec\nabla^2 \v0 u^\mu u^\nu \del_\mu \del_\nu - \Box \,.
\end{eqnarray}
\end{subequations}
If one wants to write the equations obeyed by the electromagnetic
potentials in a general frame of reference, one needs to use the
invariant operators shown in \Eqn{invops} instead of using $\del_t$
and $\vec\nabla$.  In order to complete the task, we also need to
remember that we can write
\begin{subequations}
  \label{invphiA}
\begin{eqnarray}
  \phi &\v0& u_\mu A^\mu \,, \\
  \{ 0, \vec A \} &\v0& A^\mu - (u \cdot A) u^\mu = (\eta^{\mu\nu} -
  u^\mu u^\nu) A_\nu \,,
  \label{invA}
\end{eqnarray}
\end{subequations}
where $\eta_{\mu\nu}$ is the metric, $\mathop{\rm diag}(+1,-1,-1,-1)$.
Similarly, the charge density and current density in the rest frame of
the medium can also be written in a manner that would apply to all
frames:
\begin{subequations}
  \label{invrhoJ}
\begin{eqnarray}
  c\rho &\v0& u_\mu J^\mu \,, \\
  \{ 0, \vec J \} &\v0& J^\mu - (u \cdot J) u^\mu = (\eta^{\mu\nu} -
  u^\mu u^\nu) J_\nu \,.
  \label{invJ}
\end{eqnarray}
\end{subequations}
Thus, if one wants to write \Eqn{medeq1} in its full covariant glory,
it should read like this:
\begin{eqnarray}
  \epsilon \Big( (\epsilon\mu - 1) u^\mu u^\nu \del_\mu \del_\nu +
  \Box \Big) u_\lambda A^\lambda &=& \frac1c u_\lambda J^\lambda \,.
  \label{coveq1}
\end{eqnarray}
This is clearly covariant, and \3Eqs{invops} {invphiA} {invrhoJ} can
be used to show that it reduces to \Eqn{medeq1} in the rest frame of
the medium.  Another equation can be written, using \Eqs{invA} {invJ},
that will reduce to \Eqn{medeq2} in the rest frame of the medium.
Because \Eqn{medeq} can be written in these covariant forms, my
prescription \cite{mypaper} does not ``fail'', as Franklin concluded.

It is of course true that \Eqn{coveq1} does not appear in my original
paper \cite{mypaper}.  Neither does its partner, an equation that I
omitted here as well.  The reason is that \Eqn{coveq1} looks clumsy.
The other equation, which would reduce to \Eqn{medeq2}, is clumsier.
Partly to stay away from this clumsiness, I avoided using the
potentials while writing the field equations.  Instead, in the
original paper \cite{mypaper}, I wrote the equation as
\begin{eqnarray}
  \del_\mu G^{\mu\nu} = \frac1c J^\nu \,,
  \label{dG}
\end{eqnarray}
where $J^\nu$ is the free current density 4-vector, and
\begin{eqnarray}
  G^{\mu\nu} = \epsilon (E^\mu u^\nu - E^\nu u^\mu) +
  \frac1\mu \varepsilon^{\mu\nu\lambda\rho} B_\lambda u_\rho \,.
  \label{G}
\end{eqnarray}
If one substitutes the expressions for $E^\mu$ and $B^\mu$ in terms of
the potential 4-vector $A^\mu$ given in my original paper, it can be
easily seen that if both sides of \Eqn{dG} are contracted with
$u_\nu$, one obtains \Eqn{coveq1}.  Similarly, by contracting both
sides of \Eqn{dG} with $(\eta_{\mu\nu}-u_\mu u_\nu)$, one obtains the
other equation that reduces to \Eqn{medeq2} in the rest frame of the
medium.  While performing this exercise, one needs to remember that
\Eqn{medeq} is valid only if the Lorenz gauge condition has been
employed, which reads
\begin{eqnarray}
  \vec\nabla \cdot \vec A + {\epsilon\mu \over c} {\del\phi \over \del
  t} = 0
\end{eqnarray}
in the medium rest frame.  In a general frame, it should be written in
the form
\begin{eqnarray}
  \del_\mu A^\mu = (1 - \epsilon\mu) (u \cdot \del) (u \cdot A) \,.
\end{eqnarray}

One final comment.  Franklin seems to argue that $\{ \phi, \vec A \}$
is a 4-vector because it satisfies a covariant equation such as
\Eqn{boxA}.  This line of argument is not safe.  The inhomogeneous
field equations, written in terms of the potentials, are covariant
{\em only} if the potentials are assumed to satisfy a covariant gauge
condition.  In non-covariant gauges such as the Coulomb gauge or the
axial gauge, the potentials do not obey a covariant set of equations,
as can be verified from any textbook on classical electromagnetic
theory.  The foursome $\{ \phi, \vec A \}$ is said to form a 4-vector
because of their relation with the electric and magnetic fields.  The
field strengths $\vec E$ and $\vec B$ are gauge invariant.  That is
why I used them in my original paper rather than the potentials, and
that is why the equations given in my paper, e.g., \Eqn{dG}, look much
neater than the same equations written by using the potentials.


\begin{thebibliography}{[9]}\itemsep=0pt

\bibitem{comment} J. Franklin: {\em Comment on `Covariant formulation of
electrodynamics in isotropic media'}. submitted to Eur. J. Phys.

\bibitem{mypaper} P. B. Pal: {\em Covariant formulation of
  electrodynamics in isotropic media}, Eur. J. Phys. 43 (2021) 015204.


\end{thebibliography}
\end{document}